# Raman spectroscopic investigations of carbon nanowalls

R. Kar[*1], N.N. Patel[2], S. Sinha[1]

[1]Laser & Plasma Surface Processing Section, Bhabha Atomic Research Centre, Mumbai-400085, India
[2] High Pressure & Synchrotron Radiation Physics Division, Bhabha Atomic Research Centre, Mumbai-400085, India

[*] Corresponding author: rajibkar.ph@gmail.com

**Abstract:** Carbon nanowall (CNW) is a two dimensional graphitic nanostructures with many potential applications like field emission, catalyst support for fuel cell and negative electrode for batteries. CNWs were grown here on Si substrates by plasma enhanced chemical vapor deposition using $N_2/CH_4$ plasma at $650^0C$ temperature. Deposited CNWs were characterized by scanning electron microscopy and Raman spectroscopy. Results obtained by Raman spectroscopic techniques are extensively studied. It shows existence of D, G, $D^/$, $G^/$, and $D+D^/$ bands. These bands are analyzed to obtain detail structural information about CNWs. Analysis of each Raman band is done using a Lorentzian fit and their significance have been explained from the perspectives of CNWs. Different bands appeared in Raman spectra of CNWs have further been compared with of the same bands of graphene, single-wall and multi-wall nanotubes. Similarities and differences of CNWs with these carbon nanostructures have been pointed out during analysis. Our detailed analysis of the experimental results shows that CNWs have close similarity to turbostratic graphite i.e. with substantial amount of defects present in the graphene sheets stacked over one another.

## 1. Introduction:

Carbon nanowall (CNW) is a two-dimensional graphitic nanostructure which stands on the substrate like a wall. It was first discovered by Wu et al. [1] in 2002 during carbon nanotube (CNT) growth experiment by Microwave plasma enhanced chemical vapor deposition (MPECVD) [1]. Till today CNW have been prepared by various methods including MPECVD [1-3], radio-frequency plasma-enhanced chemical vapor deposition (rf-PECVD) [4], dc plasma-enhanced chemical vapor deposition (dc-PECVD) [5, 6], hot-wire chemical vapor deposition (HWCVD) [7, 8] and electron-cyclotron resonance chemical vapor deposition (ECR-CVD) [9, 10].

CNW is a relatively new discovered nanostructure of carbon compared to more famous CNT and therefore it is also relatively less explored compared to CNT and other graphitic nanostructures. Extensive research is currently going on this material. It is shown that CNW can be used as a good catalyst support structure for fuel cell [11] and it can also be used as the negative electrode for lithium ion battery [12]. Moreover, because of its high surface area and high aspect ratio, the field-emission capabilities of CNWs are also comparable to that of CNT [7, 13].

Raman spectroscopy has historically played an important role in the study and characterization of graphitic materials being widely used over the last four decades to characterize pyrolytic graphite, carbon fibers, glassy carbon, nanographite ribbons, [14] fullerenes, CNTs, [15] and graphene [16]. Over the years, there has been a remarkable use of Raman spectroscopy for studying $sp^2$ nanocarbons [15-17]. There are excellent reviews which use the power of Raman spectroscopic technique to reveal the structure and the electronic and phonon properties of CNTs, graphene and graphite [15, 16]. There are articles explaining differences in the Raman spectra of single, bilayer, multilayer graphene and turbostratic graphite [16, 17] and articles describing differences among graphene with single walled carbon nanotube (SWCNT) and multiwalled carbon nanotube (MWCNT). But unfortunately there are only two literatures available till date [18, 19] which investigates CNWs from the Raman spectroscopic point of view. In both the cases authors' had done considerable Raman spectroscopic investigations of their samples. But they did not give a general picture by indicating the similarities and differences of CNWs with other $sp^2$ Carbon nanostructures like graphene and CNTs. This kind of literature will always be helpful and this is where this article is primarily aimed at. Here the samples chosen for Raman spectroscopic investigations were deposited in our laboratory by a novel bias independent method [20]. But to ensure a general picture comparison of the data will be made with Raman spectroscopic results of CNWs from other workers and also with other $sp^2$ Carbon nanostructures like graphene and CNTs. Finally, a general understanding about CNWs will be formulated from the Raman spectroscopic point of view.

**2. Experimental:**

Depositions of CNWs were done by 2.45 GHz ECR plasma CVD. Figure 1 show schematic of the facility used for the synthesis of CNWs. Details of this facility is given elsewhere [21, 22].

n-type Si was chosen as substrate for the growth of CNWs and Ni was chosen as the catalyst for deposition. Two different kinds of substrates were prepared for the growth. One batch of Si substrates was annealed in air at $650^0$ C for 2 hours prior to the deposition of Ni layer while other batch was left untreated. Heating in air leads to the formation of a native oxide layer on it. A native Oxide layer on Si prior to Ni deposition helps in formation of a diffusion-barrier layer of $SiO_2$ preventing Silicide formation which is a common problem for transition metal layer deposition on Si [23, 24]. However, there are also reports of Carbon nanostructure growth without any diffusion barrier layer and sometimes even without any catalyst layer at all [2]. This kind of scheme was chosen to see the feasibility of growth of Carbon nanostructures on differently conditioned substrate in a same experiment.

15 nm thick Ni layer was deposited on them by DC magnetron sputtering. The substrates were then treated in $H_2$ plasma for 20 minutes to make Ni nanoparticles from Ni thin film. Finally they were exposed to $N_2$: $CH_4$ (1: 9) plasma for 20 minutes for the growth of CNWs. Substrate temperature and microwave power were kept fixed at $650^0$ C and 780 watt respectively throughout the experiment. After deposition it was found that thick black film is deposited on both the substrates. The films were characterized for further analysis.

**3. Characterization results and Discussion:**

**3.1 Scanning electron microscopy (SEM):** Figure 2 shows the SEM image of deposited film (CNW) on the air-heated Si substrate while Figure 3 shows the image of CNWs grown on unheated substrate. From these figures it is evident that the problem of Silicide formation does not occur in the unheated Si substrate and in both the cases growth of CNW is observed.

**3.2 Raman Spectroscopy:** Micro-Raman spectra of CNWs were taken by 532 nm Nd: YAG laser. Raman spectra of both heated and unheated sample is shown separately here. Each of the obtained Raman band is deconvoluted and analyzed with Lorentzian fittings. The fitted spectra are shown in red colour and detailed de convolutions are shown in green in all figures. Detailed analysis and discussion about the spectra are done in section 4 in terms of first and second order Raman scattering.

**3.2.1 Raman spectra of the heated sample**: Figure 4 shows Raman spectrum of the heated sample between 900-1800cm$^{-1}$. The spectrum shows existence of first order G band and second order D and G$^{/}$ bands. Figure 5 shows the Raman spectrum in the wave number range from 2500-3400 cm$^{-1}$. This spectrum shows all the 2$^{nd}$ order Raman bands including G$^{/}$, D+D$^{/}$ and 2D$^{/}$. Table 1 gives the information about the peak positions and full width half maxima (FWHM) of these peaks.

**3.2.1 Raman spectra of the unheated sample:** Figure 6 shows Raman spectrum of the unheated sample between 900-1800cm$^{-1}$ showing existence of D, G and D$^{/}$ bands While Figure 7 shows the Raman spectrum between 2800-3300 cm$^{-1}$ showing G$^{/}$, D+D$^{/}$ and 2D$^{/}$ Raman bands. Figures 6 and 7 are quite similar to figures 4 and 5 respectively. The only difference remain in the case of unheated sample is the uncharacteristic presence of a radial breathing mode (RBM) as seen in figure 8. Table 2 gives the information about the peak positions and FWHM of these peaks.

Raman spectroscopic results show that there are no variations between heated and unheated samples except the presence of RBM in the unheated samples. In general, both the samples have D, G, D$^{/}$, G$^{/}$, D+D$^{/}$, 2D$^{/}$ as Raman bands while RBM is only present in the unheated sample. The further detail discussions about these bands along with their comparison with other carbon nanostructures like graphene and carbon nanotube are given in following section.

**4. Discussion on Raman spectroscopic result:**

**4.1 First order Raman bands:** From Figures 4, 6 and tables 1, 2, the band around 1590 cm$^{-1}$ is called G band and it is the only band which originates from the first order Raman scattering. It originates at the 1$^{st}$ Brillouin zone centre (Γ point) of the of graphene unit cell. At the zone centre in plane transverse optic (iTO) and longitudinal optic (LO) modes are doubly degenerate ($E_{2g}$ symmetry) and it gives rise to this band at around 1580-1590 cm$^{-1}$ [16]. G band peak is very sharp and of highest intensity intense for bulk highly ordered pyrolytic graphite (HOPG), while for a typical monolayer graphene this peak is sharp but of lower intensity compared to second order G$^{/}$ peak (will be discussed subsequently). Also, this peak has a small shoulder peak called D$^{/}$ with much lesser intensity than G peak. D$^{/}$ occurs due to second order Raman scattering and its details will be discussed later. For SWCNT, G band can be fitted normally with two Lorentzian peaks (called G$^{+}$ and G$^{-}$) due to phonon wave vector confinement along the CNT circumference and also due to the symmetry breaking that result due to nanotube diameter. For SWCNTs, G$^{+}$ and G$^{-}$ band also gives information whether the nanotube is metallic or semiconducting [15]. For MWCNT, G$^{+}$ and G$^{-}$ splitting is both small in intensity and smeared out due to the effect of the diameter distribution within the individual

MWCNTs. G-band feature in MWCNTs predominantly exhibits a weakly asymmetric characteristic lineshape, with a peak appearing close to ~1580 cm$^{-1}$.

From Figures 4 and 6 it is seen that for CNWs, G and D$^/$ are almost of equal intensities and they look like a double peak in both the cases though D$^/$ has a lesser FWHM compared to G (from Table 1and 2). The nature of G and D$^/$ observed here is the same which has been observed over the years in CNW spectra in different literatures [3, 18, 19, 25]. However, there was never enough discussion in this regard. In all of these literatures [3, 18, 19, 25] G and D$^/$ peaks were of almost equal intensities. The reasons for two peaks G and D$^/$ of being comparable intensity for CNWs can be understood more clearly after following discussions on the 2nd order Raman modes.

**4.2 Second order Raman bands:** Except G band all the other bands present in the graphitic nanostructure Raman spectra is due to second order Raman scattering. The first one present in Figure 4 and 6 are around 1350 cm$^{-1}$ and it is called D band. This band originates from the disorder induced mode in graphite of the same name [15]. This band usually arises due to the presence of in-plane substitutional heteroatoms, grain boundaries, vacancies and other defects. It originates at K point of Brilluoin zone involving one iTO phonon and one defect state [16]. This band primarily indicates defects and disappears for HOPG. D$^/$ is another second order Raman band that appears as a shoulder peak of G band and is visible around 1620 cm$^{-1}$. This peak is observed in some disordered graphitic carbon structures [18] including graphene [16]. This peak is not possible to occur under defect free conditions. It also originates due to second order Raman scattering at the K point involving one iTO phonon and one defect state. The differences between D and D$^/$ is while the former occurs due to intervalley process, the later is due to intra-valley process. This peak is not present in pure SWCNTs [15]. Comparing with other carbon nanostructures it can be said that this band is normally absent in tubular (like CNTs) or particle (like carbon nanoparticles) like graphitic nanostructures but it is always present where planar graphitic nanostructure is more directly involved in the growth (E.g.: graphene nanoribbons, CNW). This peak is observed when there are more chances of defect generation in graphene sheets. D and D$^/$ are dispersive and their position in the Raman spectra changes with incident laser wavelength [26, 27]. The dispersive behavior of D and D$^/$ bands can be understood in terms of double resonant Raman scattering [26].

Figure 5 and 7 show three distinct the 2$^{nd}$ order Raman bands of CNWs. The most important and intense band is around 2710 cm$^{-1}$. This band is normally called G$^/$ [15-17] but some authors have marked it as 2D [18] as its peak position occurs around twice the D band frequency. However, calling it 2D may be a misnomer as no defect state is involved during the formation of this peak. This peak arises from a two-phonon, intervalley, second-order process. This peak is of dispersive nature and it is routinely observed in graphene and SWCNT Raman spectra [15-17]. For the case of monolayer graphene this peak is the most intense peak in the spectra as intervalley triple resonance also occurs at this point leading to increase in its intensity. This triple-resonance condition might explain why the G$^/$ band is more intense than G band in monolayer graphene. Park et al. [28] have shown that the large intensity of the G$^/$ band for monolayer graphene can also be explained within the context of the double and triple resonance formalism. Several authors [15-17, 28] have shown that G$^/$ band can be successfully used to find out the number of layers in a graphene sample. A monolayer graphene exhibits a single and most intense Lorentzian peak due to double and triple Raman scattering. For bilayer graphene, four Lorentzian peaks can be fitted due to the splitting of π electronic structure of graphene. With an increase in number of layers, the number of double resonance process also increases and eventually the shape converges to graphite with only two Lorentzian peaks present [16, 17]. For turbostratic graphite, in which the stacking of the graphene layers is rotationally random with respect to one another shows $G^/$ band can be fitted with a single Lorentzian peak just as monolayer graphene but with a larger FWHM. The absence of an interlayer interaction between the graphene planes makes the Raman spectra of turbostratic graphite look much like that for monolayer graphene [16]. The reason for this is that the electronic structure for turbostratic graphite can be described in the same way as monolayer graphene, but now with a broadening of the $G^/$ feature due to the relaxation of the double resonance selection rules associated with the random orientation of the graphene layers with respect to each other. For monolayer graphene, FWHM is around 24 cm$^{-1}$ unlike 45-60 cm$^{-1}$ for turbostratic graphene [29]. Also the relative intensity of the $G^/$ feature to that of the G-band is much smaller for turbostratic graphite than for monolayer graphene. For SWCNTs same peak can be used to confirm whether the tube is metallic or semiconducting [15]. For the case of CNWs here, this peak is weak compared to the D and G band peak but it can be fitted with a single Lorentzian peak in both the cases. The nature of this peak is similar to the $G^/$ band observed in the turbostratic graphite.

The peak around 2965 cm$^{-1}$ observed here is D+$D^/$ peak which is generally associated with the damaged graphene [17]. This is perhaps the first case where this peak is observed in CNWs. Ni et al. [18] observed a D+G peak for CNWs at 2937 cm$^{-1}$ which is not reported again still now for CNWs.

Another weak band is observed around 3247 cm$^{-1}$ for the both cases. This band seems to be arising due to the overtone of $D^/$ peak. Observation of this peak is a new feature reported for the Raman spectroscopy of CNWs.

Among all the previous literatures on CNW [1-3, 18, 19, 25], only one literature has described the complete Raman spectra of it with $G^/$ band [18]. Other authors did not mention whether this peak is present or absent as their interest mainly lied on the confirmation of the presence of graphitic nanostructures and they confirmed it from presence of D and G bands. From the Raman spectra of CNWs presented here and from the Raman spectra of Ni et al. [18] it can be seen that $G^/$ peak for CNW is much less intense than G peak. So, any chance of triple resonance can be safely neglected. Again $G^/$ band in a CNW spectrum fits with a single Lorentzian like monolayer graphene and turbostratic graphite but a larger FWHM like makes it closer relative to turbostratic graphite. Since, CNW is formed due to stacking of graphene sheets without having any favorable orientations this is reflected in its turbostratic nature. The presence of D+$D^/$ peak in CNW like damaged graphene here further strengthens this observation.

**4.3 RBM:** The most unusual feature in the Raman spectra of CNWs is the presence of RBM shown in Figure 8. It shows an uncharacteristically sharp peak at around 185.63 cm$^{-1}$ for the unheated sample. Radial breathing mode is normally taken as a signature of SWCNT formation and one can calculate the tube diameter from this peak [15]. Conventionally RBM generates due to the coherent vibration of C atoms in the radial direction and SWCNT diameter of nanotube can be found by using a relation as $\omega_{RBM} = \dfrac{A}{d_t} + B$ ; where A and B are constants and $d_t$ is the nanotube diameter [15]. There are many articles which calculated values of A and B experimentally [30-33]. From Dresselhaus et al. [15], if SWCNT diameter remains between 1.5 ± 0.2 nm, then values of A and B can be taken as 234 cm$^{-1}$ and 10 respectively. This data gives a nanotube diameter value of 1.33 nm in this case. But it is very unlikely for RBM to appear for CNWs. In fact, presence of RBM is in the Raman spectra of CNWs is reported only once till now by Wu et al. [1]. They found the presence of breathing mode in CNW Raman spectra and suggested that occurrence of the breathing mode is due to the existence of some hollow cavity inside the flakes of CNWs. After that to the best of our knowledge RBM is not seen in CNW Raman spectra till now. So, RBM cannot be treated as a general Raman spectrum feature of CNWs as it did not appear in both the cases. Rather, in the lines of Wu et al. it can be said that RBM only appears in CNW when hollow tubular deposition happens with it due to lesser amount of process control in the plasma [1].

**5. Summary and Conclusion:**
Here we have presented an overview of the Raman spectroscopy of CNWs in an extensive manner. The basic theoretical considerations have been taken into account to explain the obtained Raman spectra of CNWs. Different peaks have been identified and compared with graphene, graphite and CNT. Structural similarity of CNWs with turbostratic graphite and damaged graphene is pointed out on the basis of Raman spectroscopic data. The variations of intensities among G, D$^/$, and G$^/$ peaks have been explained from the perspective of CNWs. From the information provided by Raman it is concluded that CNWs grow when multilayer graphene sheets stacked on one another with random orientations which is explained from the characteristics of G$^/$ peak. Intense D, D$^/$ peak and presence of D+D$^/$ peak conclude that these graphene sheets forming CNWs do have defects and it is a common feature for CNWs. Uncharacteristic presence of RBM is observed for one sample and it is suggested that this cannot be a regular feature of CNW spectra. Its occurrence is related to hollow tubular deposition in plasma.

**Table 1.** Details of different Raman bands of air heated sample.

| Sample ID | Name of the band | Position of the peaks (cm$^{-1}$) | FWHM |
|---|---|---|---|
| Heated | D | 1352.84 | 40.58 |
| | G | 1591.59 | 38.56 |
| | D$^/$ | 1621.20 | 20.97 |
| | G$^/$ | 2716.12 | 69.27 |
| | D+D$^/$ | 2967.22 | 83.87 |
| | 2D$^/$ | 3247.24 | 65.69 |

**Table 2.** Details of different Raman bands of unheated sample.

| Sample ID | Name of the band | Position of the peaks (cm$^{-1}$) | FWHM |
|---|---|---|---|
| Unheated | RBM | 185.63 | 13.12 |
| | D | 1351.68 | 46.47 |
| | G | 1593.15 | 38.53 |
| | D$^/$ | 1619.92 | 19.93 |
| | G$^/$ | 2712.8 | 67.99 |
| | D+D$^/$ | 2965.65 | 88.39 |
| | 2D$^/$ | 3238.42 | 63.34 |

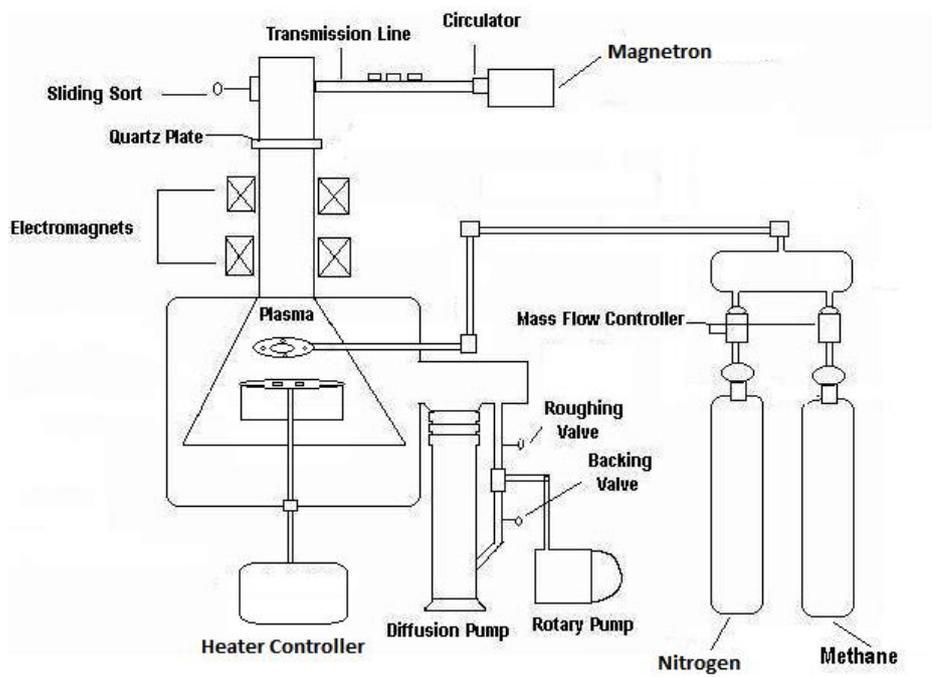

**Figure 1: Schematic of the ECR-CVD system used for CNW growth**

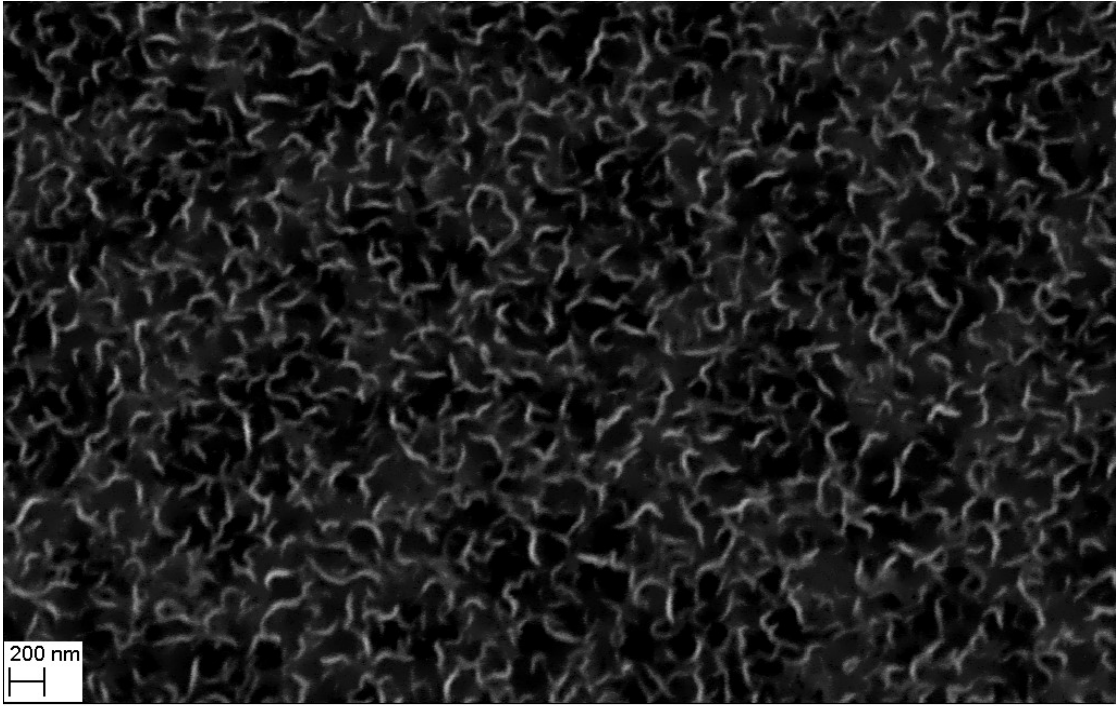

**Figure 2:** Carbon Nanowalls grwon on air heated Si Substrate

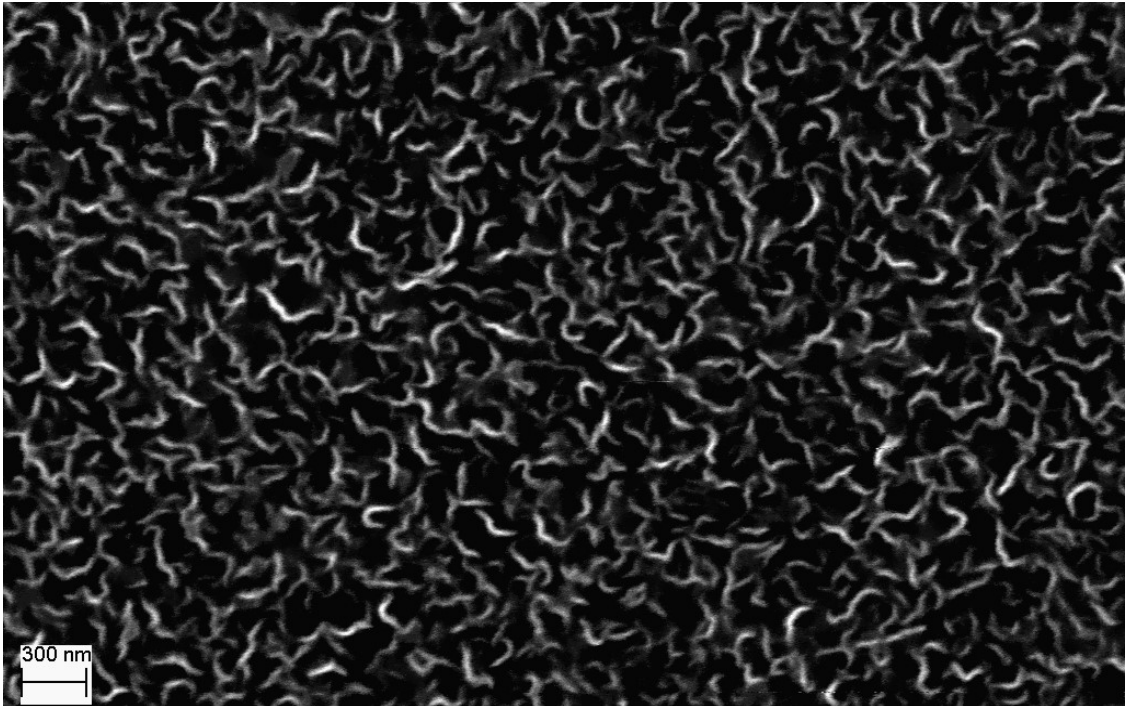

**Figure 3: Carbon Nanowalls grown on unheated substrate**

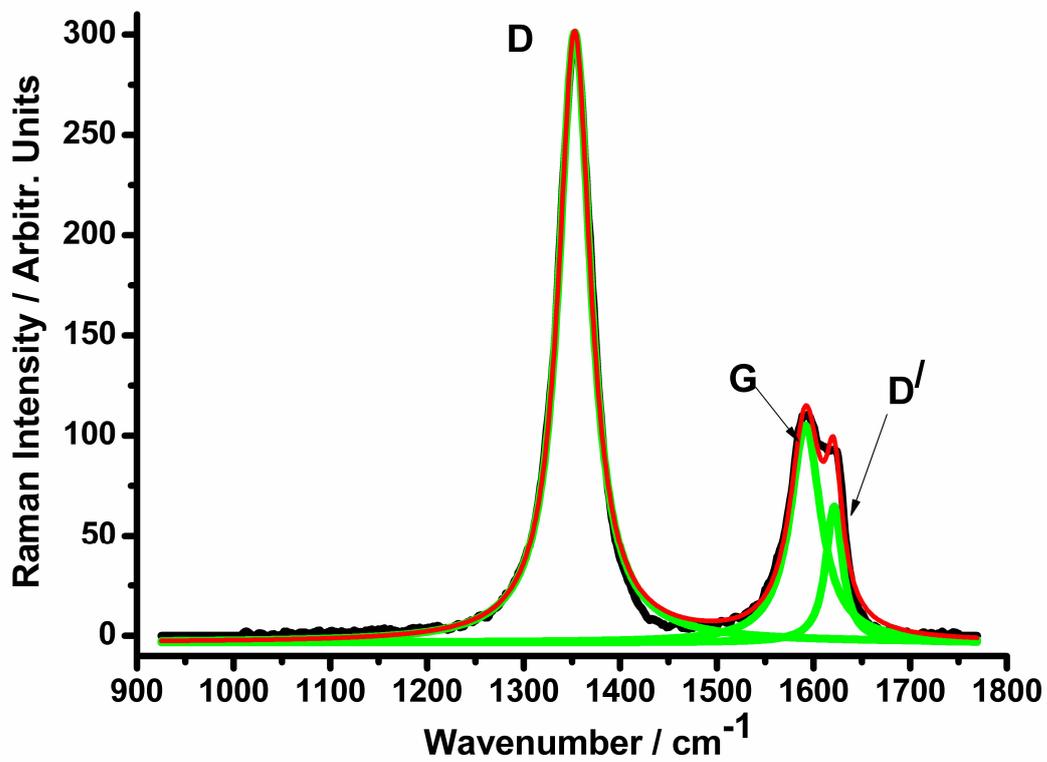

Figure 4: Raman spectra of CNWs grown on air heated substrate showing Raman shifts obtained between 900-1800 cm$^{-1}$

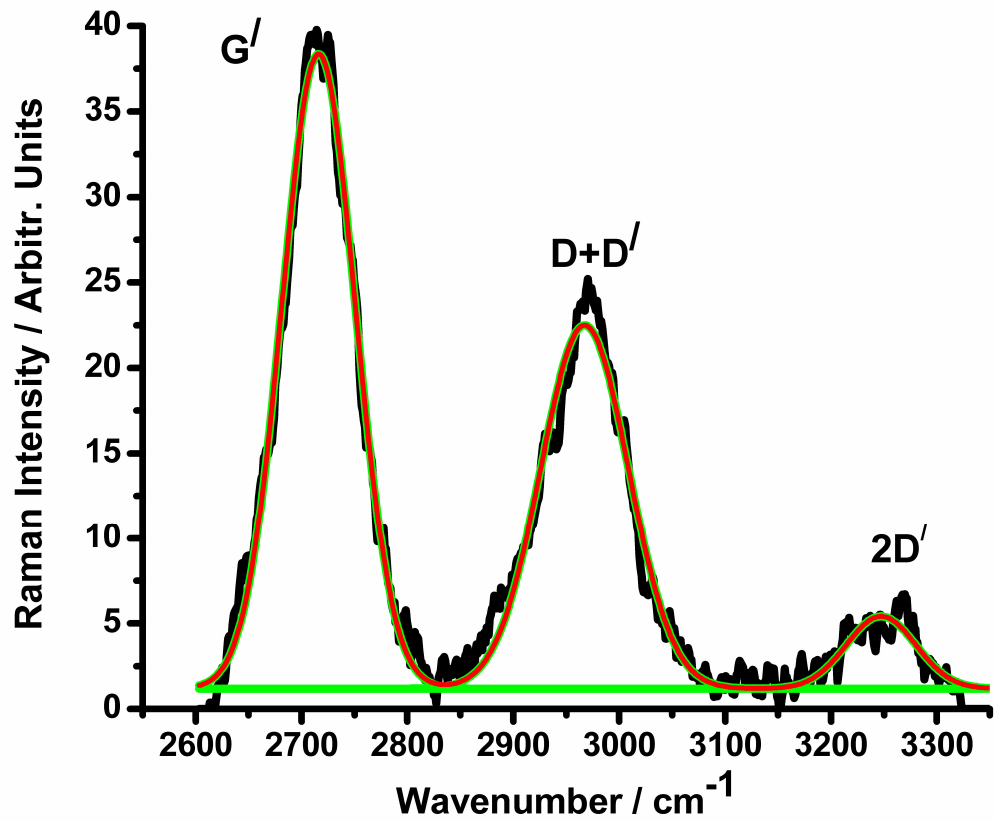

Figure 5. Raman spectra of CNWs grown on air heated substrate showing Raman shifts obtained between 2500-3400 cm$^{-1}$.

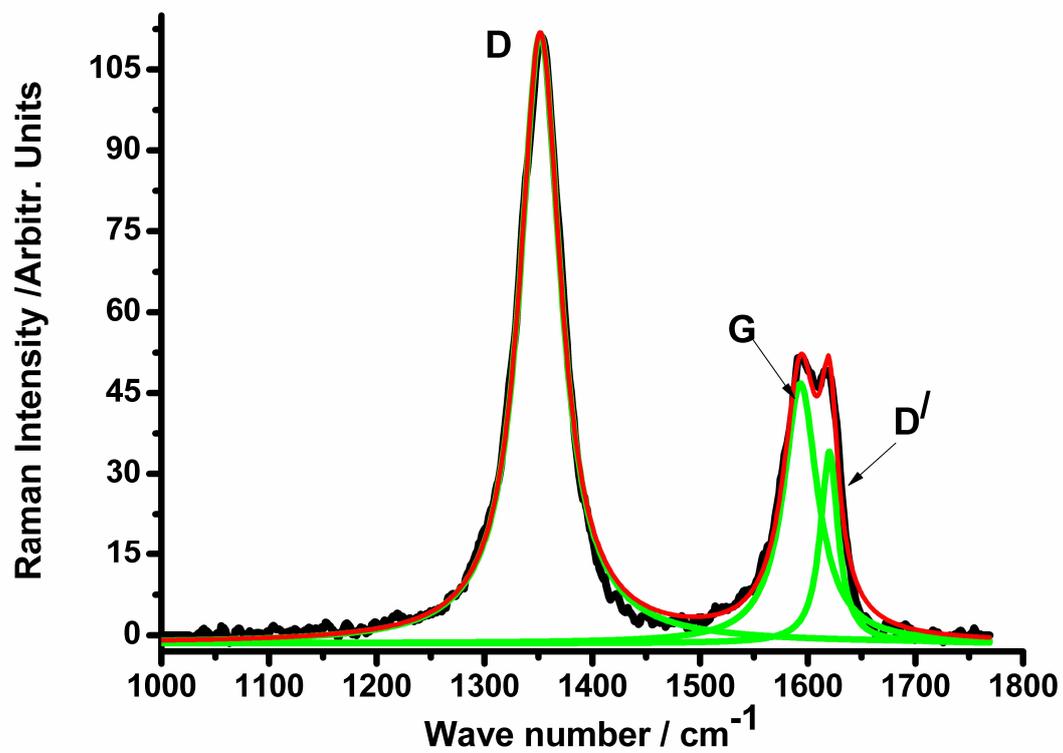

Figure 6. Raman spectra of CNWs grown on unheated substrate showing Raman shifts between 1000-1800 cm$^{-1}$.

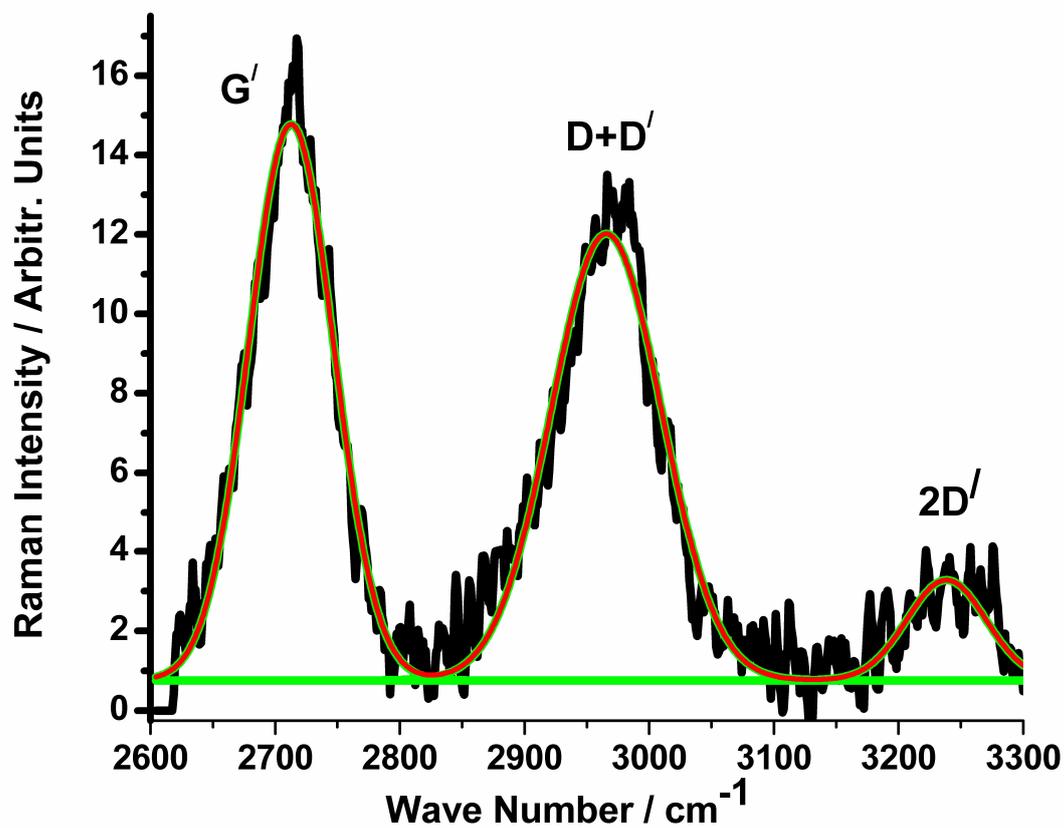

Figure 7. Raman spectra of CNWs grown on unheated substrate showing Raman shifts obtained between 2800-3300 cm$^{-1}$.

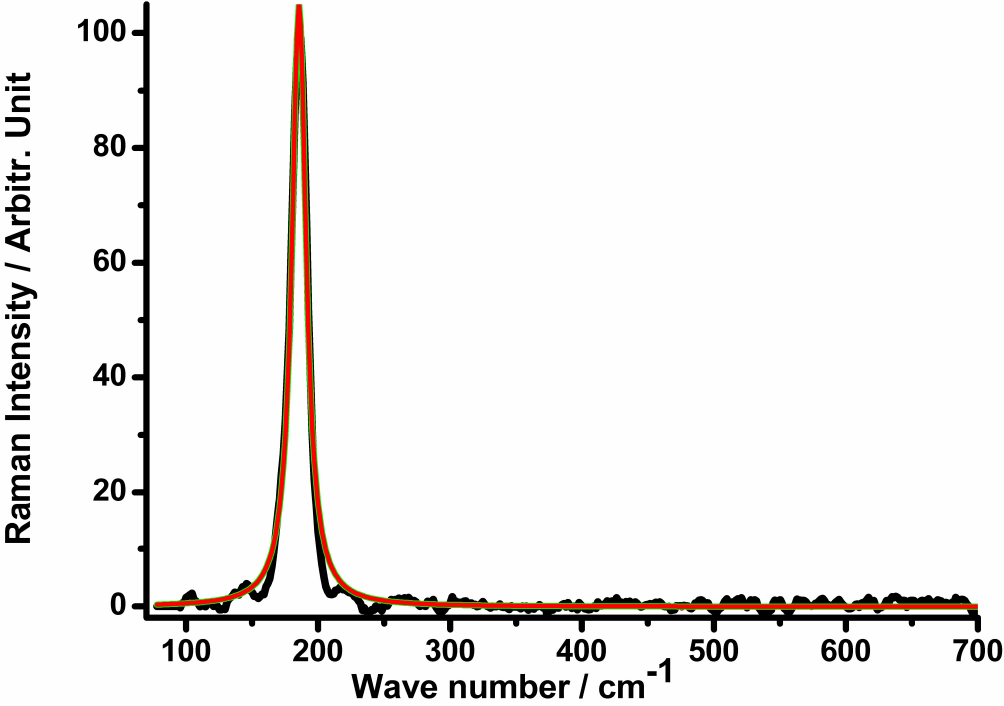

Figure 8. Raman spectra of CNWs grown on unheated substrate showing Raman shifts obtained between 100-700 cm-1 showing existence of RBM.